\documentclass[aps,pra,superscriptaddress,showpacs,showkeys,a4paper,superscriptaddress,10pt,notitlepage,twocolumn]{revtex4-1}

\usepackage[utf8]{inputenc}
\usepackage[english]{babel}
\usepackage{amsmath}
\usepackage{amssymb}
\usepackage{amsfonts}
\usepackage{graphicx}
\usepackage{bm}
\usepackage{cancel}
\usepackage{upgreek}
\usepackage{mathtools}
\usepackage[T1]{fontenc}

\usepackage{longtable}
\usepackage{cellspace}
\usepackage[tiny,center,uppercase]{titlesec}
\usepackage{longtable}

\usepackage{xcolor}
\definecolor{link_blue}{RGB}{52,46,157}
\usepackage[colorlinks,citecolor=link_blue,linkcolor=link_blue,urlcolor=link_blue]{hyperref}

\renewcommand{\vec}{\boldsymbol}

\newcommand{\opA}[1]{\boldsymbol{\mathsf{#1}}}
\newcommand{\opa}[1]{\mathsf{#1}}
\newcommand\ri{\mathrm{i}}

\DeclareMathOperator{\re}{Re}

\begin{document}
	
\title{Analytic model of a multi-electron atom}

\author{O.\ D.\ Skoromnik}
\email[]{olegskor@gmail.com}
\affiliation{Max Planck Institute for Nuclear Physics, Saupfercheckweg 1, 69117 Heidelberg, Germany}
\author{I.\ D.\ Feranchuk}
\email[Corresponding author: ]{ilya.feranchuk@tdt.edu.vn}
\affiliation{Atomic Molecular and Optical Physics Research Group, Ton Duc Thang University, 19 Nguyen Huu Tho Str., Tan Phong Ward, District 7, Ho Chi Minh City, Vietnam}
\affiliation{Faculty of Applied Sciences, Ton Duc Thang University, 19 Nguyen Huu Tho Str., Tan Phong Ward, District 7, Ho Chi Minh City, Vietnam}
\affiliation{Belarusian State University, 4 Nezavisimosty Ave., 220030, Minsk, Belarus}
\author{A.\ U.\ Leonau}
\affiliation{Belarusian State University, 4 Nezavisimosty Ave., 220030, Minsk, Belarus}
\author{C.\ H.\ Keitel}
\affiliation{Max Planck Institute for Nuclear Physics, Saupfercheckweg 1, 69117 Heidelberg, Germany}

\begin{abstract}
  A fully analytical approximation for the observable characteristics of many-electron atoms is developed via a complete and orthonormal hydrogen-like basis with a single-effective charge parameter for all electrons of a given atom. The basis completeness allows us to employ the secondary-quantized representation for the construction of regular perturbation theory, which includes in a natural way correlation effects, converges fast and enables an effective calculation of the subsequent corrections. The hydrogen-like basis set provides a possibility to perform all summations over intermediate states in closed form, including both the discrete and continuous spectra. This is achieved with the help of the decomposition of the multi-particle Green function in a convolution of single-electronic Coulomb Green functions. We demonstrate that our fully analytical zeroth-order approximation describes the whole spectrum of the system, provides accuracy, which is independent of the number of electrons and is important for applications where the Thomas-Fermi model is still utilized. In addition already in second-order perturbation theory our results become comparable with those via a multi-configuration Hartree-Fock approach.
\end{abstract}

\maketitle

The choice of the initial approximation for the single-electron wave functions (SEWF), plays an important role in modern quantum chemistry, both in the framework of the density functional theory or the solution of the Schr\"{o}dinger equation. It is well known that the numerical solutions of the Hartree-Fock (HF) equations \cite{HartreeA1928wave-2,*HartreeA1928wave-1,*FockA1930naeherungs} provide the best possible zeroth-order approximation for SEWF. The inclusion of many HF configurations (MCHF) or application of various post-HF methods \cite{doi:10.1021/cr2001383,RevModPhys.79.291,Rev.Comp.Chem.5.65.1994,PhysRevA.92.062501,PhysRevA.94.062508,PhysRev.46.618,Hirschfelder1964255} allows one to include corrections due to correlation effects.

However, despite the great efficiency of modern numerical algorithms \cite{grant2007relativistic,fischer1977hartree}, simple analytical approximations \cite{doi:10.1063/1.455064,*doi:10.1063/1.460447,jensen2007introduction,szabo1989modern,RevModPhys.23.69} still play an important role for many applications, where there is no need for extremely high accuracy, but a simple algorithm of repeated calculations of atomic characteristics is required. For example, the models based on, e.g., the Thomas-Fermi \cite{Thomas_1927,*FermiA1926medoto} or multi-parametric screening hydrogen \cite{hau2012high} approximations are widely used in computational plasma \cite{PhysRevLett.109.065002,Scott2001689,Chung20053,LEE1987131,Stewart1966} and X-ray physics \cite{hau2012high,Toraya:to5147}, crystallography \cite{Feranchuk:we0009,Waasmaier:sh0059,Toraya:to5147} or semiconductors physics \cite{PhysRevB.89.035306,SchulzeHalberg2013323,PhysRevB.57.6286}. In addition, the simplest possible inclusion of screening corrections in various cross sections like bremsstrahlung \cite{SELTZER198595} or pair production \cite{PhysRev.93.768,PhysRev.93.788} is required for later usage in particle-in-cell computer codes for simulation of strong laser-matter interaction \cite{RevModPhys.84.1177}, where computational efficiency is crucial.

In the present work we suggest a new basis set of fully analytical SEWF, which on the one hand provides a sufficiently accurate analytical zeroth-order approximation and on the other hand allows one to construct regular perturbation theory (RPT) for the inclusion of higher-order corrections. Our basis set includes the hydrogen-like wave functions with a single-variational parameter, namely the effective charge $Z^{*}$, which is identical for all SEWF of a given atom. The fact that the effective charge is identical for all SEWF is the principal difference of our approach in comparison with the inclusion of the multi-parametric screening corrections \cite{PhysRev.36.57,1402-4896-31-6-012,Feranchuk:we0009} or the quantum defect method \cite{Seaton01101958}.

The identical effective charge for all wave functions automatically provides the complete and orthonormal basis and, consequently, renders the transition into the secondary-quantized representation natural. We have demonstrated that the analytical zeroth-order approximation contains the whole spectrum of a multi-electron atom and constructed a perturbation theory series, which converges fast with the rate $\sim 1/10$. In addition, we stress here that the accuracy of our results does not depend on the number of electrons in an atom, i.e., our approximation is uniformly available for all atoms or ions. Moreover, the results via second-order perturbation theory, are comparable with those via MCHF.

In addition, our approach is distinct from the one based on the application of the Coulomb-Sturmian basis set, which was successfully employed for the approximation of the SEWF in a variety of nonrelativistic \cite{ROTENBERG1962262,doi:10.1063/1.1742383} and relativistic scattering problems \cite{PhysRevLett.100.113201,PhysRevA.89.032712,PhysRevLett.107.093202,PhysRevA.88.062707,0953-4075-45-18-181001,PhysRevA.85.062707,PhysRevA.86.062701,0953-4075-44-8-083001}. However, as highlighted in Ref.~\cite{ROTENBERG1962262} only the ground state wave function has a direct physical meaning. Consequently, it is problematic to interpret the occupation numbers for other non-ground states and to calculate the observable characteristics of multi-electron atoms such as their densities or form factors. Furthermore, the significant advantage of the hydrogen-like basis set is the knowledge of the closed-form expression through Whittaker functions for the Coulomb Green function, which in the Coulomb-Sturmian case is represented as a sum over Sturmian wave functions \cite{0022-3700-18-17-003,doi:10.1063/1.1665081,veselov1986theory}. This analytical expression for the Coulomb Green function allowed us to perform all summations via intermediate states in perturbation theory including both the discrete and continuous spectra in closed form.

First of all, let us demonstrate the effectiveness of our basis for the calculation of the atomic ground-state energies of nonrelativistic atoms. For this purpose, we write down the Hamiltonian of an atomic system with a nucleus charge $Z$ and $N$ electrons in atomic units in the secondary-quantized representation \cite{LandauQM}
\begin{align}
  \opa H &= \opa H_{0} + \opa W,\label{eq:1}
  \\
  \opa H_{0} &= \sum_{\nu}\langle\nu|\frac{\opA p^{2}}{2}-\frac{Z^{*}}{r}|\nu\rangle \opa a^{\dag}_{\nu}\opa a_{\nu},\label{eq:2}
  \\
   \opa W &= \sum_{\nu\nu_{1}}\langle\nu|\frac{-(Z - Z^{*})}{r}|\nu_{1}\rangle \opa a^{\dag}_{\nu}\opa a_{\nu_{1}} \nonumber
  \\
  &+ \frac{1}{2}\sum_{\nu\nu_{1}\mu\mu_{1}}\langle\nu|\langle\nu_{1}|\frac{1}{|\vec r - \vec r'|}|\mu_{1}\rangle|\mu\rangle \opa a^{\dag}_{\nu}\opa a^{\dag}_{\nu_{1}}\opa a_{\mu}\opa a_{\mu_{1}}. \label{eq:3}
\end{align}
Here the greek letters represent the collective quantum number $\nu=nlmm_{s}$ (or $\vec k lmm_{s}$ for the continuous spectrum) for the hydrogen-like wave function $\varphi_{\binom{nlm}{\vec k l m}}(Z^{*}\vec r)\chi_{m_{s}}(s) = \langle\vec r|\binom{nlm}{\vec k l m}\rangle\langle s|m_{s}\rangle = \langle\vec r|\langle s|\nu\rangle$ with the effective charge $Z^{*}$ and $\langle\nu|\nu_{1}\rangle = \delta_{\nu\nu_{1}}$. The fermionic operators anticommute $\{\opa a_{\nu},\opa a^{\dag}_{\nu'}\} = \delta_{\nu\nu'}$ and by acting on the $N$ particle state create the $N+1$ particle state $|\nu\lambda_{1}\ldots\lambda_{N}\rangle = \opa a^{\dag}_{\nu}|\lambda_{1}\ldots\lambda_{N}\rangle$ \footnote{We use the notations of Ref.~\cite{FeynmanB1998statistical} for the secondary-quantized representation.}. The Hamiltonian (\ref{eq:1}) is the exact expression written in the hydrogen-like basis, since we have only added and subtracted the term $Z^{*}/r$.

\begin{figure}[t]
  \centering
  \includegraphics[width=\columnwidth]{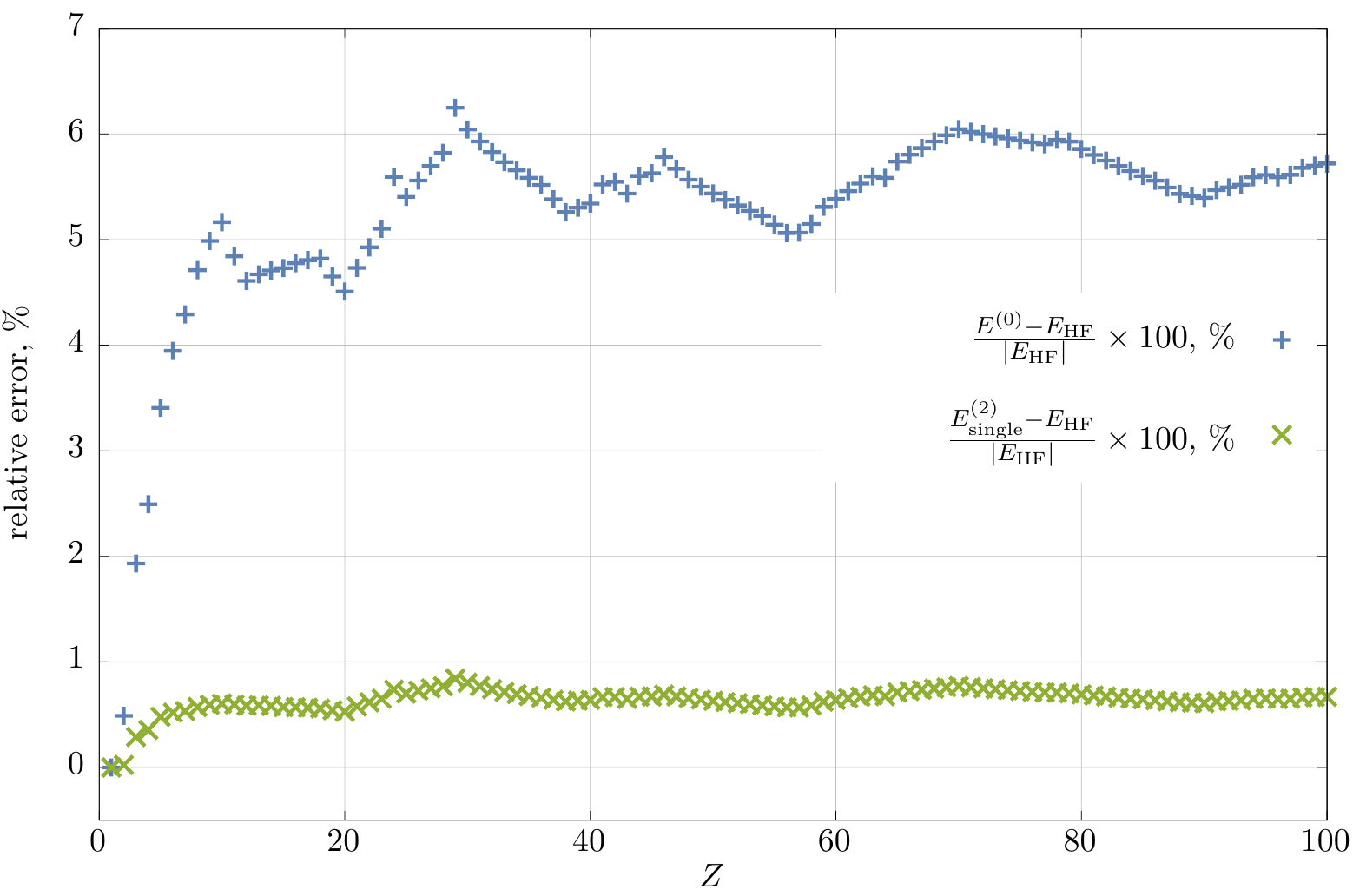}
  \caption{(color online) The relative error of the total atomic energy as a function of the nucleus' charge $Z$ (the number of electrons is equal to $Z$). The blue $+$-like crosses represent the relative error of the zeroth-order analytical energies and the green $\times$-like crosses include the single-electron excitation correction. Both cases are compared to the corresponding HF results \cite{Saito2009836,Joensson20132197}.}\label{fig:1}
\end{figure}
If the effective charge $Z^{*}$ is known, then the single-particle Hamiltonian of the zeroth-order approximation $\opa H_{0}$ is well defined. Consequently, the first question that we need to answer is how to calculate $Z^{*}$. For this we firstly performed the variational calculation, i.e., we choose the trial state vector $|\lambda_{1}\ldots\lambda_{N}\rangle$, which is characterized with a set of occupation numbers $g_{\lambda_{1}}, \ldots, g_{\lambda_{N}}$, $g_{\lambda_{k}} = 0$ or $1$ and calculated the energy of the system
\begin{align}
  E(Z^{*}) &= \langle\lambda_{1}\ldots\lambda_{N}|\opa H|\lambda_{1}\ldots\lambda_{N}\rangle\nonumber
  \\
  &= -Z^{*}(2Z-Z^{*})A + Z^{*}B,\label{eq:4}
\end{align}
where $A = \sum_{k=1}^{N} g_{\lambda_{k}}/(2n_{k}^{2})$, $B = J + K$ is the sum of the Coulomb integral $J = 1/2\int \rho(\vec r)\rho(\vec r')/|\vec r - \vec r'|d\vec rd\vec r'$ and the exchange integral $K = -1/2\int\sum_{k,l=1}^{N}g_{\lambda_{k}}g_{\lambda_{l}}\rho_{\lambda_{k}}(\vec r,\vec r')\rho_{\lambda_{l}}^{*}(\vec r,\vec r')/|\vec r - \vec r'|d\vec r d\vec r'$. In addition, $\rho(\vec r) = \sum_{k=1}^{N}g_{\lambda_{k}}\rho_{\lambda_{k}}(\vec r)$, $\rho_{\lambda_{k}}(\vec r) = |\psi_{\lambda_{k}}(\vec r)|^{2}$, $\rho_{\lambda_{k}}(\vec r, \vec r') = \psi_{\lambda_{k}}(\vec r)\psi_{\lambda_{k}}^{*}(\vec r')$ and $\psi_{\lambda_{k}}(\vec r)$ is the hydrogen wave function, i.e., $Z=1$. The quantity $B$ can be calculated analytically for an arbitrary atom (see Appendix~\ref{sec:appendixA}).
\begin{table}[t]
  \centering
  \begin{tabular}{| c | c | c | c | c |}
    \hline \hline
     & $E^{(2)}$ & $E_{\mathrm{var}}$ & $E_{\mathrm{MCHF}}$ & $E_{\mathrm{HF}}$ \\
    \hline
    $\textrm{H}^-$ & -0.532 & -0.528 & -0.528 & -0.488 \\ 
    He & -2.907 & -2.904 & -2.903 & -2.861 \\ 
    Li & -7.467 & -7.478 & -7.477 & -7.433  \\
    He $2\, ^{3}\mathrm{S}$ & -2.172 & -2.175 & -2.175 & -2.174 \\
    He $2\, ^{1}\mathrm{S}$ & -2.154 & -2.146 & -2.146 & -2.143 \\
    \hline \hline
  \end{tabular}
  \caption{The comparison of the energy in a.u. of the second-order perturbation theory in the hydrogen-like basis Eq.~(\ref{eq:19}) with the values via HF \cite{Saito2009836,doi:10.1139/v96-170,doi:10.1063/1.1701568,*cohen10.1139.65.178,*doi:10.1139/p66-263,*TrefftzA1957oscillator}, MCHF \cite{doi:10.1063/1.1701568}, variational estimation and $1/Z$ expansion ($\mathrm{H}^{-}$, He, para- and ortho-He \cite{PhysRevA.81.032118,Frolov2015,PhysRev.126.1470} and Li \cite{PhysRevA.73.022503}).}
  \label{tab:1}
\end{table}

Here we stress the extremely important fact that due to the property of the Coulomb potential the variable change $\tilde{\vec r} = Z^{*}\vec r$ leads to the simple energy dependence on the effective charge, which is given explicitly. Consequently, the analytical minimization of $E(Z^{*})$ with respect to the $Z^{*}$ yields the desired values of the effective charge and the energy of the system
\begin{align}
  Z^{*} &= Z - \frac{B}{2A}, \label{eq:5}
  \\
  E^{(0)} &= -AZ^{*2}.\label{eq:6}
\end{align}

Proceeding further, let us construct the perturbation theory due to the operator $\opa W$. For this we utilize the eigenstates of the zeroth-order Hamiltonian $\opa H_{0}$ and calculate the first correction to the energy of the system. Since $\opa H_{0}$ is exactly diagonalizable, this is trivial to perform
\begin{align}
  \Delta E^{(1)} &= \langle\lambda_{1}\ldots\lambda_{N}|\opa W|\lambda_{1}\ldots\lambda_{N}\rangle \nonumber
  \\
  &= -Z^{*}(Z-Z^{*})2A + Z^{*}B.\label{eq:7}
\end{align}

By observing Eq.~(\ref{eq:7}) we can conclude that the above value of the effective charge $Z^{*}$ (\ref{eq:5}), found from the variational estimation leads to the vanishing first-order correction to the energy of the system $\Delta E^{(1)} = 0$.

Before discussing the second-order correction let us briefly consider the accuracy of the zeroth-order approximation. Since the effective charge $Z^{*}$ is defined via Eq.~(\ref{eq:5}), the energy of the system $E^{(0)}(Z,\{g_{\lambda}\})$ in Eq.~(\ref{eq:6}) depends only on the charge of the nucleus and the set of occupation numbers $\{g_{\lambda}\}$, which should determine the minimal energy and satisfy the normalization condition $\sum_{k} g_{\lambda_{k}} = N$. We stress here that all our results, both for the zeroth-order approximation and for the second-order perturbation theory, are valid not only for atoms, but also for ions. Therefore, we write in the normalization condition $N$ and not $Z$, since in the general case $N\neq Z$.

Simple analytical calculations based on Eqs.~(\ref{eq:3}) - (\ref{eq:6}) demonstrate that the optimal choice of the occupation numbers is given according to the ``Aufbau'' or Madelung-Janet-Klechkovskii rule \cite{KlechkovskiiA1962justification,Madelung1936,doi:10.1021/ed056p714}. For example, from the two sets of the occupation numbers $[\mathrm{Ar}]4\mathrm{s}^{1}$ and $[\mathrm{Ar}]3\mathrm{d}^{1}$ for potassium ($Z=19$), the former set possesses minimal energy (compare $-571.305$ versus $-568.473$). In Fig. \ref{fig:1} and Appendix~\ref{sec:appendixC} (Table~\ref{tab:2}) the energy $E^{(0)}$ is compared with the results obtained via nonrelativistic HF equations \cite{Saito2009836}. As can be concluded from Fig.~\ref{fig:1} the chosen SEWF basis leads to a uniform approximation, i.e., it provides a relative accuracy of $\sim5\%$ with respect to HF for all elements of the periodic table, which is considerably better than the Thomas-Fermi approximation \cite{LandauQM,Thomas_1927,*FermiA1926medoto}.

Let us proceed with the calculation \footnote{The details of all calculations are presented in supplementary information.} of the correction to the energy of the system in second-order perturbation theory. Due to the two-particle structure of the perturbation potential $\opa W$ the only non-vanishing intermediate states in second-order are described by the state vectors $|\lambda_{1}\ldots\sigma_{k}\ldots\sigma_{l}\ldots \lambda_{N}\rangle$, in which $\lambda_{k}$ and $\lambda_{l}$ are replaced via intermediate states $\sigma_{k}$ and $\sigma_{l}$, respectively. Consequently, the second-order correction to the energy of the system reads
\begin{align}
  \Delta E^{(2)} = -\frac{1}{2}\sum_{k<l}\sideset{}{'}\sum_{\sigma_{k}\sigma_{l}}\frac{|W_{\lambda_{k}\lambda_{l}, \sigma_{k}\sigma_{l}}|^{2}}{E_{\sigma_{k}} + E_{\sigma_{l}} - E_{\lambda_{k}} - E_{\lambda_{l}}},\label{eq:8}
\end{align}
where $W_{\lambda_{k}\lambda_{l}, \sigma_{k}\sigma_{l}} = \langle\lambda_{1}\ldots\lambda_{N}|\opa W|\lambda_{1}\ldots\sigma_{k}\ldots\sigma_{l}\ldots \lambda_{N}\rangle$.

In this equation the sum is performed over all substitutions $\lambda_{k},\lambda_{l}$ with $\sigma_{k},\sigma_{l}$, $k,l = \{1\ldots N\}$ and the primed sum over $\sigma_{k},\sigma_{l}$ represents the sum over all possible quantum numbers excluding the ground state. It is convenient to split the total second-order correction to the energy as the sum of $\Delta E^{(2)}_{\text{single}}$, when only one electron goes into an intermediate state, and $\Delta E^{(2)}_{\text{multi}}$, when two electrons undergo the transition into intermediate states. In the first case the intermediate-state vector is $|\lambda_{1}\ldots\sigma_{k}\ldots\lambda_{N}\rangle$ with $\sigma_{k}\neq\lambda_{k}$ and in the second case $|\lambda_{1}\ldots\sigma_{k}\ldots\sigma_{l}\ldots\lambda_{N}\rangle$ with $\sigma_{k}\neq\lambda_{k}\lambda_{l}$ and $\sigma_{l}\neq\lambda_{k}\lambda_{l}$.

We continue the calculation of the single-electron excitation. The required matrix elements can be easily evaluated and represented in compact form
\begin{align}
  \langle\lambda_{1}\ldots\lambda_{N}|&\opa W|\lambda_{1}\ldots\lambda_{k-1}\sigma_{k}\lambda_{k+1}\ldots\lambda_{N}\rangle \nonumber
  \\
  &= \langle\lambda_{k}|U_{\lambda_{k}}|\sigma_{k}\rangle - \sum_{l\neq k} \langle\lambda_{l}|V_{\lambda_{k}\lambda_{l}}^{(2)}|\sigma_{k}\rangle,\label{eq:9}
\end{align}
where $U_{\lambda_{k}} = V^{(1)} + \sum_{l\neq k}V^{(2)}_{\lambda_{l}\lambda_{l}}$, $V^{(1)} = -Z^{*}(Z-Z^{*})/r$, $V^{(2)} = Z^{*}/|\vec r - \vec r'|$ and $V^{(2)}_{\lambda_{k}\lambda_{l}} = \langle\lambda_{k}|V^{(2)}|\lambda_{l}\rangle$. Here we also carried out the variable change $\vec r \to Z^{*}\vec r$ in order to separate out the explicit dependence on $Z^{*}$.

\begin{figure*}[t]
  \centering
  \includegraphics[width=\columnwidth]{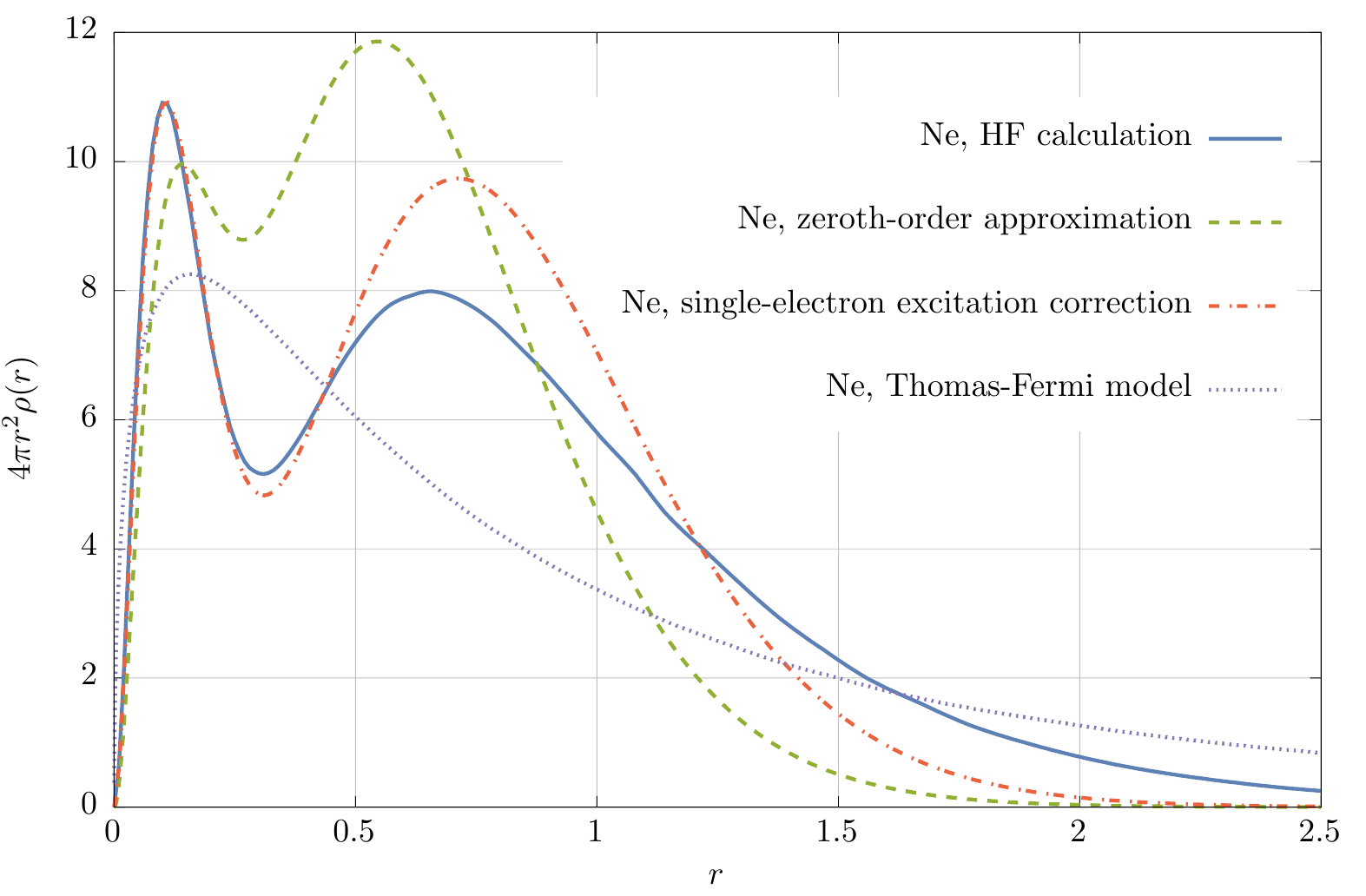}
  \includegraphics[width=\columnwidth]{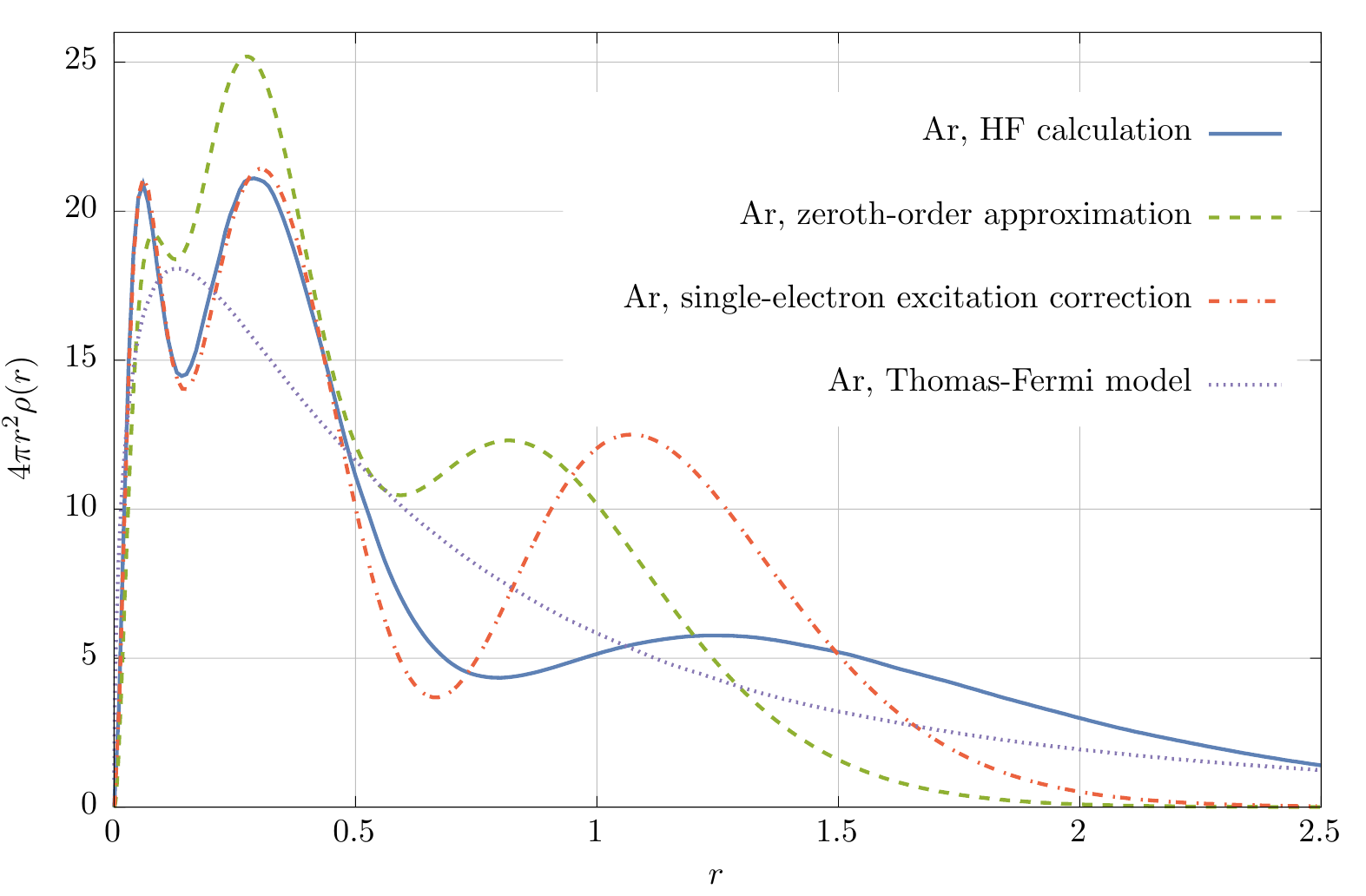}
  \caption{(color online) The dependence of the radial density Eq.~(\ref{eq:20}) for Ne and Ar atoms on the radial coordinate $r$. The blue solid line is HF calculation \cite{Joensson20132197,PhysRevA.16.891}, the green dashed line is an analytical zeroth-order approximation, the red dashed-dotted line includes a single-electron excitation first-order correction and the purple dotted line is Thomas-Fermi model \cite{LandauQM}.}\label{fig:2}
\end{figure*}
Since the required matrix element has been evaluated, we have to perform the summation over intermediate states. Here we face one of the largest advantages of using a hydrogen-like basis set as the required summation can be performed in closed form \cite{Dalgarno70,PhysRev.178.126,Dalgarno282}, since the single-particle Coulomb Green function is well known \cite{jmp.1.524039,PhysRevLett.10.469,jmp.1.1704153,veselov1986theory} and is expressed through the decomposition over spherical harmonics $Y_{lm}(\Omega)$ and Whittaker functions \cite{whittaker1996course} $W_{\kappa,\mu}(r,r'),M_{\kappa,\mu}(r,r')$
\begin{widetext}
  \begin{align}
    G_{ZE} (\vec r, \vec r') &= \sum_{\sigma_{i}}\frac{\langle\vec r|\sigma_{i}\rangle\langle\sigma_{i}|\vec r'\rangle}{E_{\sigma_{i}} - E}=\sum_{lm} \frac{1}{rr'} G_{ZEl} ( r, r')Y^*_{lm}(\Omega) Y_{lm}(\Omega'),\label{eq:10}
    \\
    G_{ZEl} (r,r')&= \frac{\nu}{Z} \frac{\Gamma (l + 1 - \nu)}{\Gamma(2l + 2)} M_{\nu, l + 1/2} \left(\frac{2Z}{\nu} r_{<} \right)  W_{\nu, l+1/2} \left(\frac{2Z}{\nu} r_{>} \right),\label{eq:11}
  \end{align}
  where $\nu = \frac{Z}{\sqrt{-2E}}$, $r_> = \max(r,r')$ and $r_< =\min(r,r')$.
\end{widetext}

In order to calculate $\Delta E^{(2)}_{\text{single}}$ one needs to take into account that in Eq.~(\ref{eq:8}) not all states are present. Moreover, since we are dealing with a multi-electron system we have to take into account the Pauli exclusion principle. It reveals itself here in the subtractions of the occupied states from the Green function of the electrons with the same spin as the electron that undergoes the transition into intermediate states. Consequently, we introduce the reduced Coulomb Green function
\begin{align}
  \tilde{G}_{E_{\lambda_{k}}-\ri\delta}^{\lambda_{k}} = G_{E_{\lambda_{k}}-\ri\delta} - \sum_{l}{}^{*} \frac{|\lambda_{l}\rangle\langle\lambda_{l}|}{E_{\lambda_{l}} - (E_{\lambda_{k}} - \ri\delta)},\label{eq:12}
\end{align}
where $\lambda_{k}$ in $\tilde{G}^{\lambda_{k}}_{E_{\lambda_{k}}-\ri\delta}$ and the starred sum describe the subtractions with the same spin as the electron $\lambda_{k}$. With the help of Eqs.~(\ref{eq:8})--(\ref{eq:12}) we obtain the energy of the system including the single-electron excitation correction
\begin{align}
  E^{(2)}_{\text{single}} &= E^{(0)} + \Delta E^{(2)}_{\mathrm{single}} = E^{(0)}  + \sum_{k=1}^{N}\Delta E^{(2)}_{\text{single},\lambda_{k}},\label{eq:13}
\end{align}
where
\begin{align}
  \Delta E^{(2)}_{\text{single},\lambda_{k}} = &-\langle\lambda_{k}|U_{\lambda_{k}}\tilde{G}_{E_{\lambda_{k}}-\ri\delta}^{\lambda_{k}}U_{\lambda_{k}}|\lambda_{k}\rangle \nonumber
  \\
  &+ 2\re\sum_{r\neq k} \langle\lambda_{k}| U_{\lambda_{k}} \tilde{G}_{E_{\lambda_{k}}-\ri\delta}^{\lambda_{k}} V_{\lambda_{k}\lambda_{r}}^{(2)*} |\lambda_{r}\rangle \nonumber
  \\
  &- \sum_{\substack{l\neq k\\r\neq k}} \langle\lambda_{l}| V_{\lambda_{k}\lambda_{l}}^{(2)} \tilde{G}_{E_{\lambda_{k}}-\ri\delta}^{\lambda_{k}}V_{\lambda_{k}\lambda_{r}}^{(2)*} |\lambda_{r}\rangle.\label{eq:14}
\end{align}
Here we also pay attention to the fact that both the Green function and the states $|\lambda_{k,l}\rangle$ in Eq.~(\ref{eq:14}) are referred to the hydrogen wave functions, since due to the variable change $\vec r \to Z^{*}\vec r'$ the dependence on $Z^{*}$ is given explicitly and the $Z^{*}$ in the denominator of Eq.~(\ref{eq:10}) is cancelled with the one from the matrix element.

The results obtained via Eqs.~(\ref{eq:13}), (\ref{eq:14}) are presented in Fig.~\ref{fig:1} and in Appendix~\ref{sec:appendixC} (Table~\ref{tab:2}) for nuclear charges $Z=1..100$, however, for $Z$ larger than $20$ the incorporation of relativistic corrections would be required for precise values. One can observe that the inclusion of the single-electron excitation reduces the relative error by one order of magnitude. The inclusion of $\Delta E^{(2)}_{\text{single}}$ does not take into account correlation effects and consequently the corrected wave functions still remain in the class of SEWF. For this reason the condition $|E^{(2)}_{\text{single}}|<|E_{\mathrm{HF}}|$ holds as the solution of HF equations realizes a minimum of the functional. In addition, single-electron correction in third-order perturbation theory should also not be taken into account, since its value is smaller than the corresponding second-order correlation correction.

Let us proceed with the calculation of the correlation correction $\Delta E^{(2)}_{\mathrm{multi}}$. The required matrix elements in this case are represented as
\begin{align}
  \langle\lambda_{1}\ldots\lambda_{N}|&\opa W|\lambda_{1}\ldots\sigma_{k}\ldots\sigma_{l}\ldots \lambda_{N}\rangle \label{eq:15}
  \\
  &=\langle\lambda_{k}|\langle\lambda_{l}|V^{(2)}|\sigma_{k}\rangle|\sigma_{l}\rangle - \langle\lambda_{l}|\langle\lambda_{k}|V^{(2)}|\sigma_{k}\rangle|\sigma_{l}\rangle. \nonumber
\end{align}

Here only the two-particle part of the operator $\opa W$ contributes, since both $\sigma_{k}$ and $\sigma_{l}$ can not be equal to both $\lambda_{k}$ and $\lambda_{l}$, respectively.

In order to perform a summation over intermediate states in this case we need to obtain a two-particle Coulomb Green function, which is not known. However, with the help of the identity 
\begin{align}
  \int_{-\infty}^{\infty}\frac{dt}{(t + a - \ri \delta/2)(t - b + \ri \delta/2)} &= -\frac{ 2\pi \ri}{a + b - \ri\delta}\label{eq:16}
\end{align}
the two-particle Green function can be represented as a convolution of two single-particle ones
\begin{align}
  G_{E - \ri\delta}(1,2) = -\int_{-\infty}^{\infty}\frac{dt}{2\pi\ri}G_{t + \frac{E - \ri\delta}{2}}\otimes G_{-t + \frac{E - \ri\delta}{2}}.\label{eq:17}
\end{align}

The only nontrivial operation remained is to perform the required subtractions, taking into account the Pauli exclusion principle. Let us illustrate this in the lithium case. For example, for the matrix element $\langle\lambda_{1}\lambda_{2}\lambda_{3}|\opa W|\lambda_{1}\sigma_{1}\sigma_{2}\rangle$, $\sigma_{1}$ can not be equal to $\lambda_{1}\lambda_{2}\lambda_{3}$. The same applies to $\sigma_{2}$ \footnote{The situation when $\sigma_{2} = \lambda_{3}$ and $\sigma_{1}\neq\lambda_{1}\lambda_{2}\lambda_{3}$ or equivalently $\sigma_{1} = \lambda_{2}$ and $\sigma_{2}\neq\lambda_{1}\lambda_{2}\lambda_{3}$ is taken into account in the single-electron excitation.}. Consequently, taking into account the spin orthogonality, we need to subtract $\frac{|\lambda_{2}\rangle\langle\lambda_{2}|}{(t+ E_{0}/2 - \ri\delta/2)- E_{\lambda_{2}}}$ from the Green function with the index $t$, with $E_{0} = E_{\lambda_{2}}+ E_{\lambda_{3}}$. The Green function with the index $-t$ undergoes two subtractions, namely $\frac{|\lambda_{1}\rangle\langle\lambda_{1}|}{(-t+ E_{0}/2 - \ri\delta/2)- E_{\lambda_{1}}}$ and $\frac{|\lambda_{3}\rangle\langle\lambda_{3}|}{(-t+ E_{0}/2 - \ri\delta/2)- E_{\lambda_{3}}}$.

With the help of the above notation for the reduced Green function Eq.~(\ref{eq:12}), the correlation correction is written as
\begin{widetext}
  \begin{align}
    \Delta E^{(2)}_{\mathrm{multi}} &= \sum_{k<l}\lim_{\delta\to0}\Bigg(\int_{-\infty}^{\infty}\frac{dt}{2\pi\ri}\langle\lambda_{k}|\langle\lambda_{l}|V^{(2)}\tilde{G}_{t+(E_{\lambda_{k}}+E_{\lambda_{l}}-\ri\delta)/2}^{\lambda_{k}}\otimes \tilde{G}_{-t+(E_{\lambda_{k}}+E_{\lambda_{l}}-\ri\delta)/2}^{\lambda_{l}}V^{(2)}|\lambda_{k}\rangle|\lambda_{l}\rangle \label{eq:18}
    \\
    &\mspace{150mu}-\re\int_{-\infty}^{\infty}\frac{dt}{2\pi\ri}\langle\lambda_{k}|\langle\lambda_{l}|V^{(2)}\tilde{G}_{t+(E_{\lambda_{k}}+E_{\lambda_{l}}-\ri\delta)/2}^{\lambda_{k}}\otimes \tilde{G}_{-t+(E_{\lambda_{k}}+E_{\lambda_{l}}-\ri\delta)/2}^{\lambda_{l}}V^{(2)}|\lambda_{l}\rangle|\lambda_{k}\rangle\delta_{m_{s\lambda_{k}}m_{s\lambda_{l}}}\Bigg), \nonumber
  \end{align}
\end{widetext}
which is valid for an arbitrary atom or ion due to the pairwise character of the correlation contribution.

Finally, combining all together we obtain the total energy of the system in second-order perturbation theory
\begin{align}
  E^{(2)} = E^{(0)} + \Delta E^{(2)}_{\mathrm{single}} + \Delta E^{(2)}_{\mathrm{multi}}. \label{eq:19}
\end{align}

In order to demonstrate the effectiveness of our basis set we have evaluated the energy Eq.~(\ref{eq:19}) in second-order perturbation theory for the ground states of $\mathrm{H}^{-}$, He and Li and the excited states for He, namely ortho- and para-helium (see Table.~\ref{tab:1}). We note here, that due to the degeneracy of ortho- and para-helium the perturbation theory should be modified, i.e., the zeroth-order state is defined as \cite{Feranchuk20112550} $|\psi^{2\, ^{3}S, 2\, ^{1}S}\rangle = \frac{1}{\sqrt{2}} (|\lambda_{1}\uparrow,\lambda_{2}\downarrow\rangle\pm|\lambda_{1}\downarrow,\lambda_{2}\uparrow\rangle)$, the so called perturbation theory for the doubly degenerate energy levels \cite{LandauQM} (See also supplementary material). In addition, for some energy levels in Table~\ref{tab:1} ($\mathrm{H}^{-}$, He, He $2\, ^{1}\mathrm{S}$) their energy values within a complete second-order perturbation theory are smaller than the corresponding exact results via the variational method. This is related to the fact that in our calculations we employ perturbation theory series, which convergence to the exact value can be oscillatory in some problems \cite{Feranchuk2015,Feranchuk1995370,0305-4470-19-9-030}, i.e., the absolute value of the difference between the exact and the approximate results is decreasing in each order of perturbation theory, however, in the second order the approximate value is smaller than the exact result, while in the third order it is larger respectively. The mathematical proof of this convergence property requires additional investigations.

At last we want to demonstrate that our basis set provides a good approximation not only for the integral characteristics of the system but also for the local ones. For this we have evaluated the radial electron density $4\pi r^{2}\rho(r)$, with the wave function
\begin{align*}
  |\psi^{(1)}\rangle &= |\lambda_{1}\ldots\lambda_{N}\rangle + \sum_{i=1}^{N}\sideset{}{'}\sum_{\sigma_{i}}\frac{W_{\sigma_{i}\lambda_{i}}|\lambda_{1}\ldots \sigma_{i}\ldots \lambda_{N}\rangle}{E_{\lambda_{i}} - E_{\sigma_{i}} },
\end{align*}
which includes the first-order single-electron excitation correction over $\opa W$. Consequently, one finds the expectation value of the density operator $\langle\psi^{(1)}|\uprho(\vec r)|\psi^{(1)}\rangle = \langle\psi^{(1)}|\sum_{\nu\nu'}\psi^{\dag}_{\nu}(\vec r)\psi_{\nu'}(\vec r)\opa a^{\dag}_{\nu}\opa a_{\nu'}|\psi^{(1)}\rangle$ up to first-order in $\opa W$:
\begin{widetext}
  \begin{align}
    \langle\psi^{(1)}|\uprho(Z^{*}\vec r)|\psi^{(1)}\rangle &= \sum_{i=1}^{N}\varphi^{\dag}_{\lambda_{i}}(Z^{*}\vec r)\varphi_{\lambda_{i}}(Z^{*}\vec r) \label{eq:20}
    \\
    &\mspace{-90mu}- 2Z^{*2}\re\sum_{i=1}^{N}\varphi^{\dag}_{\lambda_{i}}(Z^{*}\vec r)\left[\int d\vec r' \tilde{G}^{\lambda_{i}}_{E_{\lambda_{i}} - \ri \delta}(Z^{*}\vec r, r')U_{\lambda_{i}}(\vec r')\psi_{\lambda_{i}}(\vec r') - \sum_{\substack{l=1\\l\neq i}}^{N}\int d\vec r'\tilde{G}^{\lambda_{i}}_{E_{\lambda_{i}} - \ri \delta}(Z^{*}\vec r, r')V^{(2)}_{\lambda_{l}\lambda_{i}}(\vec r')\psi_{\lambda_{l}}(\vec r')\right]. \nonumber
  \end{align}
\end{widetext}

The dependence of the density on the radial variable $r$ for Ne and Ar, which possess spherically symmetric radial density, is presented in Fig.~\ref{fig:2}. We immediately observe that already in the fully analytical zeroth-order approximation the error in the density does not exceed $\sim 20\%$ in comparison with the corresponding HF value \cite{PhysRevA.16.891}. Moreover, our fully analytical result provides much better agreement than the quasi-classical Thomas-Fermi model \cite{LandauQM,Thomas_1927,*FermiA1926medoto}. At the same time, the inclusion of the single-electron excitation correction improves the agreement with HF significantly.
\begin{figure*}[t]
  \centering
  \includegraphics[width=\columnwidth]{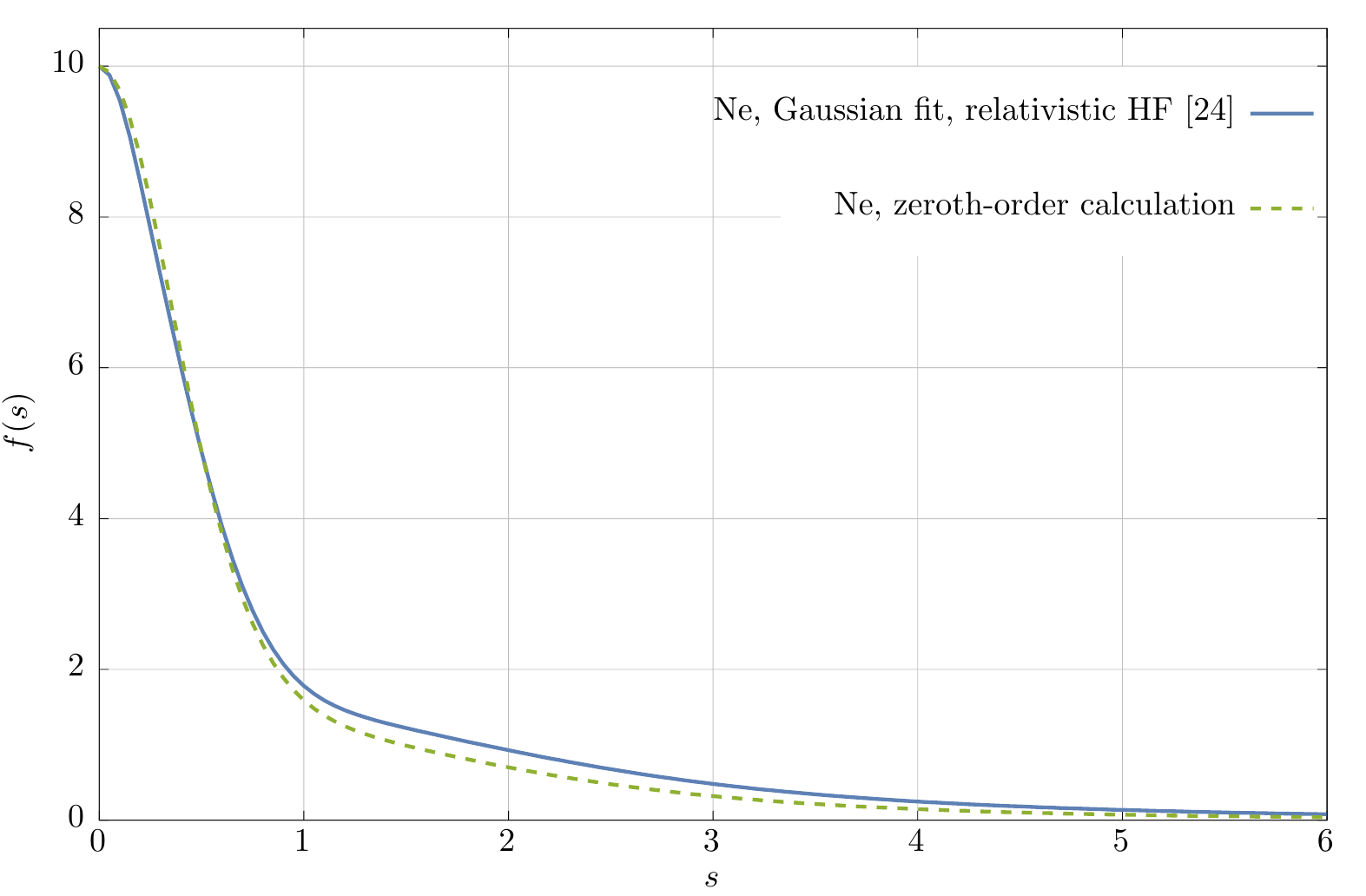}
  \includegraphics[width=\columnwidth]{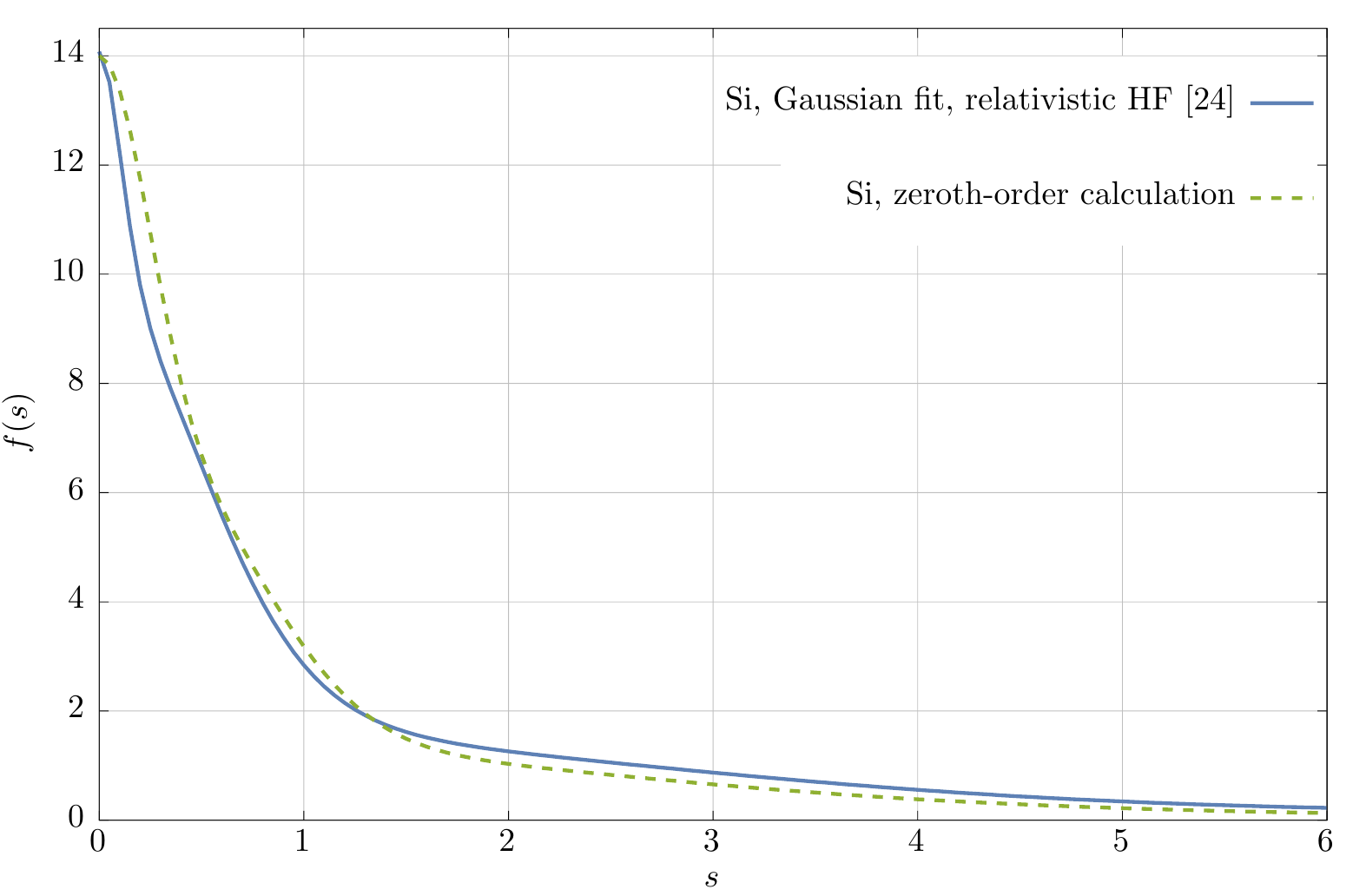}
  \\
  \includegraphics[width=\columnwidth]{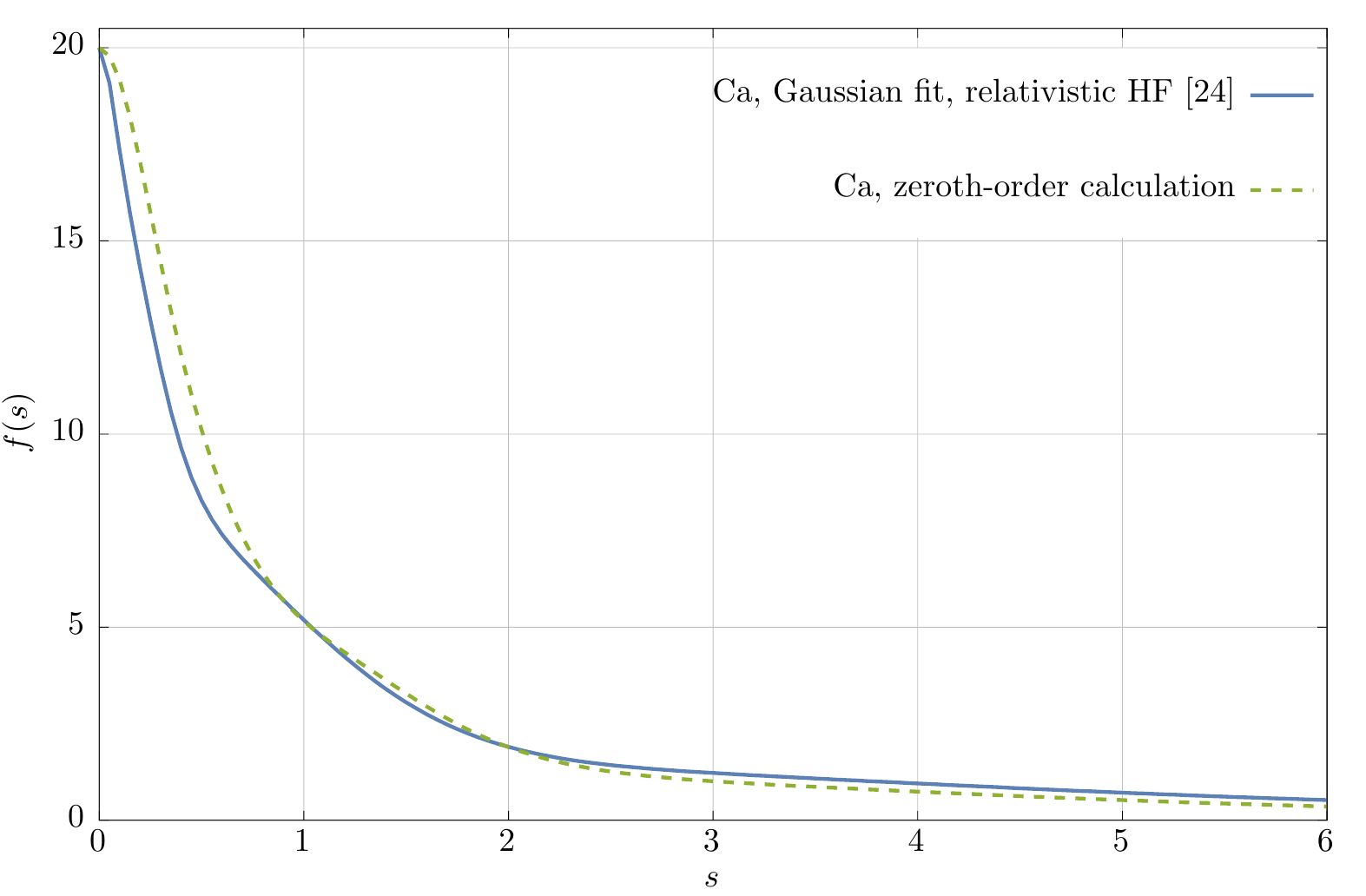}
  \includegraphics[width=\columnwidth]{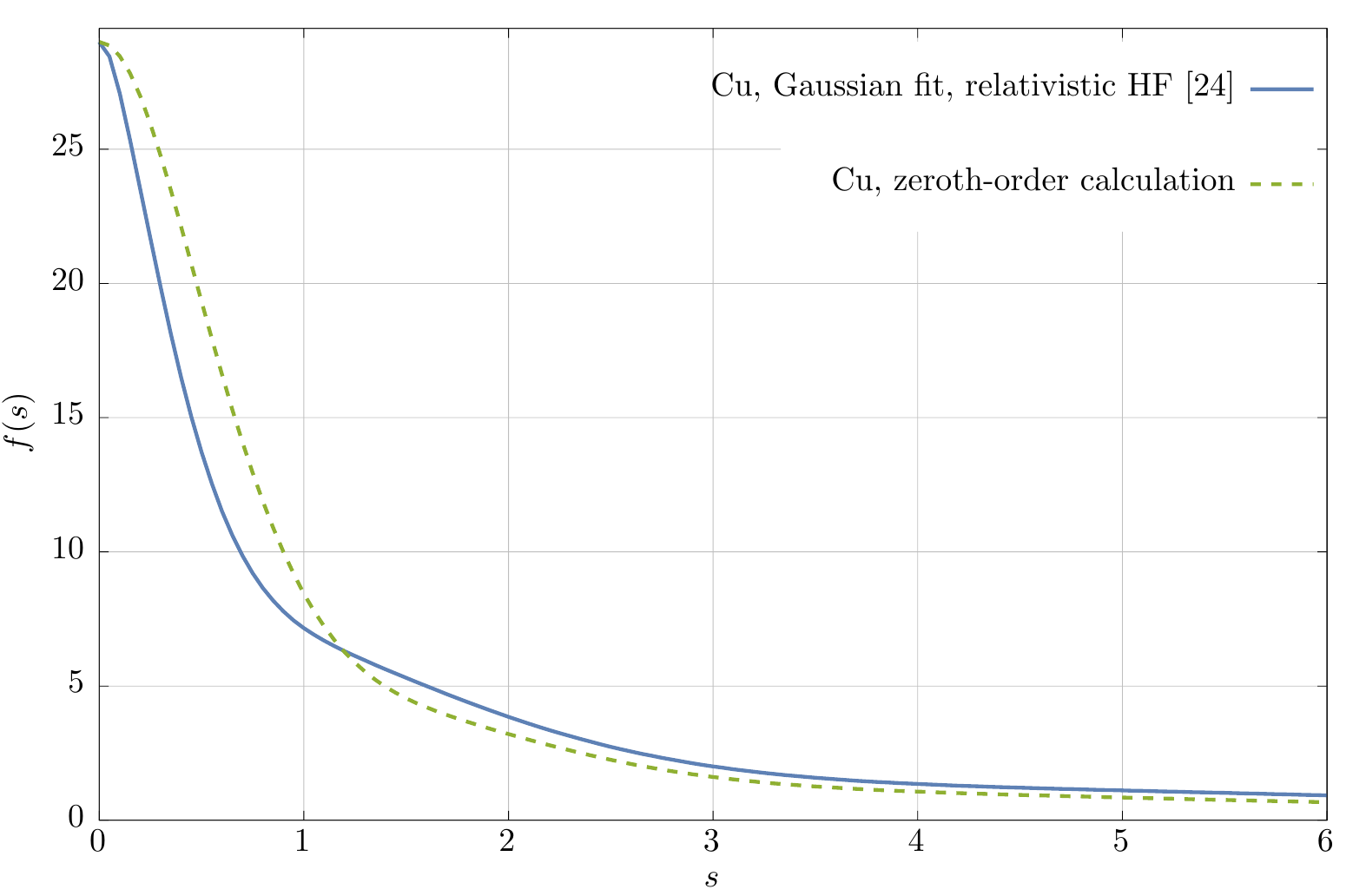}
  \caption{(color online) The dependence of the atomic scattering factors for Ne, Si, Ca and Cu atoms on the parameter $s = \sin\theta/\lambda\,[\text{\AA}^{-1}]$ . The blue solid line is a gaussian fit taken from \cite{Waasmaier:sh0059}, the green dashed line is our analytical zeroth-order approximation.}\label{fig:3}
\end{figure*}

As was mentioned in the introduction, when a large number of repeated calculations need to be performed, the simplest possible expressions for the electronic density and the spectrum of the system are required. This happens in, e.g., computer codes for plasma simulations like \texttt{CRETIN} \cite{Scott2001689}, \texttt{FLYCHK} \cite{Chung20053}, \texttt{LASNEX-DCA} \cite{LEE1987131} where Thomas-Fermi model is used for the determination of the electrostatic potential \cite{Stewart1966} or for the calculation of X-ray scattering factors \cite{hau2012high} in crystallography, where fits of an electronic density are employed. Since our zeroth-order approximation is fully analytical it can be perfectly used for these purposes. For example, for the density of Ne, the atom with the largest number of electrons in the first row of the periodic table, one obtains in the zeroth-order approximation
\begin{align}
  \rho_{\mathrm{Ne}}^{(0)} = \frac{Z^{*3}}{8\pi}e^{-2Z^{*}r}(16 + e^{Z^{*}r}(2+Z^{*}r(Z^{*}r - 2))),\label{eq:21}
\end{align}
which is extremely simple and provides better qualitative agreement with HF than the Thomas-Fermi model, see Fig.~\ref{fig:2}. For larger atoms, expressions for the density look similar and have a structure of an exponential function which multiplies a polynomial, as can be found in Appendix~\ref{sec:appendixB}.

Moreover, the Fourier transform $f^{(0)}(\vec q) = \int \rho^{(0)}(\vec r)e^{\ri\vec q\cdot\vec r}d\vec r$ of an electronic density for an arbitrary atom or ion can also be calculated analytically in the zeroth-order approximation (see Appendix.~\ref{sec:appendixB}). However, specifically for the atoms with spherically symmetric density the following closed form expression can be employed \cite{benediktovitch2013theoretical}
\begin{widetext}
\begin{align}
  f^{(0)}_{\text{sph.sim}}(q) &= \sum_{nl}g_{nl} F_{nl}^{\text{sph.sim}}(Z^{*},q), \quad \xi = \frac{2Z^{*}}{n},\label{eq:22}
  \\
  F^{\text{sph.sim}}_{nl} &= -\xi^{2l+3}\frac{(n-l-1)!(n+l)!}{2n}\sum_{k,m=0}^{n-l-1}\frac{\xi^{k+m}}{(2l+k+1)!(2l+m+1)!k!m!}\frac{d^{2l+1+k+m}}{d\xi^{2l+1+k+m}}\frac{1}{\xi^{2}+q^{2}}. \label{eq:23}
\end{align}
\end{widetext}

Usually for numerical applications the fit of the form $f(s) = \sum_{i}a_{i}\exp{(-b_{i}s^{2})}$ is used for the atomic scattering factors \cite{Waasmaier:sh0059,Rez:sp0059,tablesCrystallographyC2006}. Here $s = \sin\theta/\lambda\,[\text{\AA}^{-1}]$, $\theta$ is the scattering angle and $\lambda$ the wavelength of the X-ray radiation. The parameter $s$ is related to $q$ as $4\pi s\cdot 0.529177 = q$. Consequently, in Fig.~\ref{fig:3} we compare highly accurate Gaussian fits \cite{Waasmaier:sh0059} of results from relativistic HF calculations with our simple analytical expressions. For light elements we find good agreement; however, for larger elements the discrepancy increases, indicating the importance of corrections. 

Lastly, we conclude that the proposed procedure can be straightforwardly generalized for the relativistic hydrogen-like basis set, since the analytical form of the relativistic Coulomb Green function is known \cite{doi:10.1063/1.526255,0305-4470-24-1-020,*0305-4470-24-8-022}. The analytical zeroth-order wave functions can substitute the less accurate Thomas-Fermi approximation. We also remind that our calculation scheme is valid for ions, since the number of electron is not necessary equal to $Z$, which can be useful for the statistical theory of plasma. Our approach can be easily modified to include interactions with external fields, employed for the classification of excited states and applied in time dependent problems, where the knowledge of the system's spectrum is required. In addition our results provide a uniform approximation for the observable characteristics, i.e., independent of the number electrons in an atom. Furthermore, since our zeroth-order approximation yields algebraic expressions for electronic densities and scattering factors, our results can be useful for repeated calculations in other particle-in-cell and plasma computer codes for simulation laser-matter interactions.

Finally, the authors are working toward the release of the computer software, which will incorporate relativistic corrections, will allow an automated calculation of the energies of the excited states, transition matrix elements and oscillator strengths.

\begin{acknowledgments}
  The authors are grateful to S. Cavaletto, V. Triguk and S. Bragin for useful discussions, to F. Koeck for the assistance with the MPIK cluster and to N. Oreshkina and Z. Harman for providing comments for the manuscript.
\end{acknowledgments}

\appendix

\section{Explicit expressions for the zeroth-order energy}
\label{sec:appendixA}

In this Appendix we present an analytical calculation of the zeroth-order approximation for the energy of the system.

According to the discussion before Eq.~(\ref{eq:4}), in order to calculate an effective charge $Z^{*}$ one needs to evaluate the expectation value of the Hamiltonian with a trial state vector $|\lambda_{1}\ldots\lambda_{N}\rangle$. The evaluation of the matrix elements is presented in Ref.~\cite{szabo1989modern}. As a result, the calculation of the energy is reduced to the calculation of the Coulomb and exchange integrals
\begin{align}
  J &= \frac{1}{2}\int \frac{\rho(\vec r)\rho(\vec r')}{|\vec r - \vec r'|}d\vec rd\vec r',\label{eq:24}
  \\
  K &= -\frac{1}{2}\int\sum_{k,l=1}^{N}g_{\lambda_{k}}g_{\lambda_{l}}\frac{\rho_{\lambda_{k}}(\vec r,\vec r')\rho_{\lambda_{l}}^{*}(\vec r,\vec r')}{|\vec r - \vec r'|}d\vec r d\vec r',\label{eq:25}
\end{align}
where we have introduced the diagonal and of-diagonal elements of the density matrix
\begin{align}
  \rho(\vec r) &= \sum_{k=1}^{N}g_{\lambda_{k}}\rho_{\lambda_{k}}(\vec r),\label{eq:26}
  \\
  \rho_{\lambda_{k}}(\vec r) &= |\psi_{\lambda_{k}}(\vec r)|^{2},\label{eq:27}
  \\
  \rho_{\lambda_{k}}(\vec r, \vec r') &= \psi_{\lambda_{k}}(\vec r)\psi_{\lambda_{k}}^{*}(\vec r').\label{eq:28}
\end{align}

The hydrogen wave function $\psi_{\lambda_{k}}(\vec r)$ in the expressions (\ref{eq:26})--(\ref{eq:28}) is the product of the spherical harmonic $Y_{lm}(\Omega)$ and the radial wave function
\begin{align}
  \psi_{\lambda_{k}}(\vec r) = R_{nl}(r)Y_{lm}(\Omega).\label{eq:29}
\end{align}

The actual expression for $R_{nl}(r)$ is given, for example, in \cite{LandauQM}.

In the following we will use the expansion of the Coulomb interaction between electrons over spherical harmonics \cite{LandauQM}
\begin{align}
  \frac{1}{|\vec r - \vec r'|} = \sum_{j = 0}^{\infty}\sum_{s =  -j}^{j}\frac{4\pi}{2j+1}\frac{r_{<}^{j}}{r_{>}^{j+1}} Y^{*}_{js}(\Omega') Y_{js}(\Omega),\label{eq:30}
\end{align}
with $r_> = \max(r,r')$ and $r_< =\min(r,r')$. In addition, the integration of the product of three spherical harmonics yields 3j symbols \cite{cowan1981theory}
\begin{align}
  \int &Y_{l_{1}m_{1}}(\Omega)Y_{l_{2}m_{2}}(\Omega)Y_{l_{3}m_{3}}(\Omega) d\Omega \nonumber
  \\
       &= \sqrt{\frac{(2l_{1}+1)(2l_{2}+1)(2l_{3}+1)}{4\pi}} \nonumber
  \\
       &\times\begin{pmatrix}
    l_{1} & l_{2} & l_{3}
    \\
    0 & 0 & 0
  \end{pmatrix}
            \begin{pmatrix}
              l_{1} & l_{2} & l_{3}
              \\
              m_{1} & m_{2} & m_{3}
            \end{pmatrix}.\label{eq:31}
\end{align}
Consequently, the calculation of the quantities $J$ and $K$ reduces to the computation of the two integrals and summation over occupation numbers.

We proceed with the evaluation of the Coulomb part. For this, we firstly calculate
\begin{align}
  J_{\lambda_{k}\lambda_{l}} &= \frac{1}{2} \int \frac{\rho_{\lambda_{k}}(\vec r)\rho_{\lambda_{l}}(\vec r')}{|\vec r - \vec r'|}d\vec r d\vec r' \nonumber
  \\
  &= \frac{1}{2} \int \frac{|\psi_{\lambda_{k}}(\vec r)|^{2}|\psi_{\lambda_{l}}(\vec r')|^{2}}{|\vec r - \vec r'|}d\vec r d\vec r'.\label{eq:32}
\end{align}

By plugging Eq.~(\ref{eq:29}) into Eq.~(\ref{eq:32}), using expansion of Eq.~(\ref{eq:30}) and integrating out the angular variables one obtains
\begin{align}
  J_{\lambda_{k}\lambda_{l}} = \frac{1}{2}\sum_{j=0}^{\min(2l,2l_{1})}I^{j}_{nl,n_{1}l_{1}} M_{lm,l_{1}m_{1}}^{j},\label{eq:33}
\end{align}
where
\begin{align}
  M^{j}_{lm,l_{1}m_{1}} &= (-1)^{m+m_{1}}(2l+1)(2l_{1}+1)\nonumber
  \\
  &\times\begin{pmatrix}
    l & l & j
    \\
    0 & 0 & 0
  \end{pmatrix}
            \begin{pmatrix}
              l & l & j
              \\
              m & -m & 0
            \end{pmatrix}\nonumber
  \\
  &\times\begin{pmatrix}
    l_{1} & l_{1} & j
    \\
    0 & 0 & 0
  \end{pmatrix}
            \begin{pmatrix}
              l_{1} & l_{1} & j
              \\
              m_{1} & -m_{1} & 0
            \end{pmatrix},\label{eq:34}
\end{align}
and
\begin{align}
  I^{j}_{nl,n_{1}l_{1}} &= \int_{0}^{\infty}\int_{0}^{\infty}dr dr' r^{2} r^{\prime 2}|R_{nl}(r)|^{2}|R_{n_{1}l_{1}}(r')|^{2}\frac{r_{<}^{j}}{r_{>}^{j+1}} \nonumber
  \\
                        &=\int_{0}^{\infty}dr r^{2}|R_{nl}(r)|^{2}\Bigg(\frac{1}{r^{j+1}}\int_{0}^{r}r^{\prime j+2}|R_{n_{1}l_{1}}(r')|^{2}dr' \nonumber
  \\
  &\mspace{70mu}+ r^{j}\int_{r}^{\infty}r^{\prime 1 - j}|R_{n_{1}l_{1}}(r')|^{2}dr'\Bigg).\label{eq:35}
\end{align}
In addition the quantum numbers $\lambda_{k} = nlmm_{s}$ and $\lambda_{l} = n_{1}l_{1}m_{1}m_{s1}$. Therefore, the Coulomb integral is equal to
\begin{align}
  J = \sum_{k,l = 1}^{N}g_{\lambda_{k}}g_{\lambda_{l}}J_{\lambda_{k}\lambda_{l}},\label{eq:36}
\end{align}
which can be fast evaluated using computer algebra software of a choice, e.g., \texttt{MATHEMATICA}. We would like to mention here, that the same set of integrals appears constantly and in order to speed up the evaluation it makes sense to precalculate the integrals and store the values in an array as a function of quantum numbers. Consequently, once this is performed, the evaluation of the Coulomb integral for any set of quantum numbers can be done almost instantly.

In a full analogy one can calculate the exchange integral, yielding
\begin{align}
  K = \frac{1}{2}\sum_{k,l = 1}^{N}g_{\lambda_{k}}g_{\lambda_{l}} \delta_{m_{s}m_{s1}}\sum_{j = |l - l_{1}|}^{l+l_{1}}L^{j}_{nl,n_{1}l_{1}}D^{j}_{lm,l_{1}m_{1}}, \label{eq:37}
\end{align}
where
\begin{align}
  L^{j}_{nl,n_{1}l_{1}} &= \int_{0}^{\infty}dr r^{2} R^{*}_{n_{1}l_{1}}(r)R_{nl}(r)\label{eq:38}
  \\
  &\mspace{30mu}\times\Bigg(\frac{1}{r^{j+1}}\int_{0}^{r}r^{\prime 2+j}R^{*}_{nl}(r')R_{n_{1}l_{1}}(r')dr' \nonumber
  \\
  &\mspace{70mu}+ r^{j}\int_{r}^{\infty}r^{\prime 1-j}R^{*}_{nl}(r')R_{n_{1}l_{1}}(r')dr'\Bigg) \nonumber
\end{align}
and
\begin{align}
  D^{j}_{lm,l_{1}m_{1}} &= (-1)^{l + l_{1} + j}(2l+1)(2l_{1}+1)\nonumber
  \\
  &\times\begin{pmatrix}
    l_{1} & l & j
    \\
    0 & 0 & 0
  \end{pmatrix}^{2}\begin{pmatrix}
    l_{1} & l & j
    \\
    -m_{1} & m & (m_{1} - m)
  \end{pmatrix}^{2}.\label{eq:39}
\end{align}

In addition, the above discussion about calculation efficiency of the Coulomb integrals is fully applicable for the exchange integral.

\section{Explicit expressions for the zeroth-order electronic density}
\label{sec:appendixB}

In this appendix we present the explicit expressions for the electronic density in the analytical zeroth-order approximation together with its Fourier transforms for a number of selected atoms. In the following we use the notation $Z^{*}r = u$.

He
\begin{align*}
  \rho^{(0)}_{\mathrm{He}} = \frac{2Z^{*3}}{\pi}e^{-2u}, \quad f^{(0)}_{\mathrm{He}} = \frac{32Z^{*4}}{(q^{2} + 4Z^{*2})^{2}}. \label{eq:40}
\end{align*}

C
\begin{align*}
  \rho^{(0)}_{\mathrm{C}} &= \frac{2Z^{*3}e^{-2u}}{\pi}\left(1 + \frac{e^{u}}{8}\left(1-u+\frac{11}{32}u^{2}\right)+\frac{e^{u}u^{2}\cos{2\theta}}{256}\right),
  \\
  f^{(0)}_{\mathrm{C}} &= \frac{6Z^{*4}(6q^{8}+25q^{6}Z^{*2}+30q^{4}Z^{*4}+16Z^{*8})}{(q^{2}+Z^{*2})^{4}(q^{2} + 4Z^{*2})^{2}}.
\end{align*}

O
\begin{align*}
  \rho^{(0)}_{\mathrm{O}} &= \frac{2Z^{*3}e^{-2u}}{\pi}\left(1 + \frac{e^{u}}{8}\left(1-u+\frac{13}{32}u^{2}\right)-\frac{e^{u}u^{2}\cos{2\theta}}{256}\right),
  \\
  f^{(0)}_{\mathrm{O}} &= \frac{2Z^{*4}(18q^{8}+76q^{6}Z^{*2}+99q^{4}Z^{*4}+24q^{2}Z^{*6} + 64Z^{*8})}{(q^{2}+Z^{*2})^{4}(q^{2} + 4Z^{*2})^{2}}.
\end{align*}

Ne
\begin{align*}
  \rho^{(0)}_{\mathrm{Ne}} &= \frac{2Z^{*3}e^{-2u}}{\pi}\left(1 + \frac{e^{u}}{8}\left(1-u+\frac{1}{2}u^{2}\right)\right),
  \\
  f^{(0)}_{\mathrm{Ne}} &= \frac{4Z^{*4}(9q^{8}+37q^{6}Z^{*2}+42q^{4}Z^{*4}+40Z^{*8})}{(q^{2}+Z^{*2})^{4}(q^{2} + 4Z^{*2})^{2}}.
\end{align*}

\begin{widetext}
Xe
\begin{align*}
  \rho^{(0)}_{\mathrm{Xe}} &= \frac{2 e^{-2 u} Z^{*3}}{\pi } + \frac{e^{-u} \left(u^2-2 u+2\right) Z^{*3}}{8 \pi }+\frac{2 e^{-\frac{2 u}{3}} \left(4 u^4-48 u^3+216 u^2-324 u+243\right) Z^{*3}}{6561 \pi }
  \\
  &+\frac{e^{-\frac{u}{2}} \left(19 u^6-720 u^5+10080 u^4-65280 u^3+207360 u^2-276480 u+184320\right) Z^{*3}}{5898240 \pi }
  \\
  &+\frac{2 e^{-\frac{2 u}{5}}Z^{*3}}{10986328125 \pi} \Big(12 u^8-1120 u^7+41200 u^6-765000 u^5+7668750 u^4
  \\
  &-41250000 u^3+112500000 u^2-140625000 u+87890625\Big),                             
\end{align*}
while the Fourier transform can be obtained by applying the expression (\ref{eq:23}).
\end{widetext}

For all other atoms or ions similar expressions can be obtained, yielding the product of an exponential by a polynomial functions.

\section{Calculated values of the effective charges, ground state energies and their comparison with HF}
\label{sec:appendixC}
\setlength{\tabcolsep}{10pt}
\setlength\LTcapwidth{0.94\textwidth}
\begin{longtable*}{|ccccc|ccccc|} 
  \hline \hline
    $Z$ &$Z^{*}$ & $E^{(0)}$ & $E^{(2)}_{\mathrm{single}}$ & $E_{\mathrm{HF}}$ & $Z$ & $Z^{*}$ & $E^{(0)}$ & $E^{(2)}_{\mathrm{single}}$ & $E_{\mathrm{HF}}$ \\
    \hline
    1 & 1. & -0.5 & -0.5 & -0.5 & 51 & 40.3872 & -5974. & -6274.4 & -6313.49 \\
    2 & 1.6875 & -2.8477 & -2.8610 & -2.86168 & 52 & 41.2295 & -6259.8 & -6571.53 & -6611.8 \\
    3 & 2.5454 & -7.2891 & -7.4114 & -7.43273 & 53 & 42.0706 & -6553.2 & -6876.53 & -6917.98 \\
    4 & 3.3716 & -14.2096 & -14.5212 & -14.573 & 54 & 42.9104 & -6854.3 & -7189.47 & -7232.1 \\
    5 & 4.1511 & -23.6936 & -24.4115 & -24.5291 & 55 & 43.7925 & -7165.6 & -7510.2 & -7553.9 \\
    6 & 4.9127 & -36.2016 & -37.4927 & -37.6886 & 56 & 44.6732 & -7484.4 & -7838.7 & -7883.5 \\
    7 & 5.6605 & -52.0662 & -54.1107 & -54.4009 & 57 & 45.4977 & -7804.6 & -8174.35 & -8221.1 \\
    8 & 6.3823 & -71.2844 & -74.3812 & -74.8094 & 58 & 46.2332 & -8125.8 & -8516.62 & -8566.9 \\
    9 & 7.0975 & -94.4525 & -98.8188 & -99.4093 & 59 & 46.8783 & -8447.6 & -8865.53 & -8921.2 \\
    10 & 7.8073 & -121.908 & -127.769 & -128.547 & 60 & 47.6094 & -8783.9 & -9224.48 & -9283.9 \\
    11 & 8.6561 & -154.020 & -160.894 & -161.859 & 61 & 48.3384 & -9127.99 & -9591.84 & -9655.1 \\
    12 & 9.4972 & -190.415 & -198.448 & -199.615 & 62 & 49.0657 & -9479.96 & -9967.76 & -10035.0 \\
    13 & 10.3161 & -230.579 & -240.453 & -241.877 & 63 & 49.7914 & -9839.95 & -10352.3 & -10423.5 \\
    14 & 11.1294 & -275.254 & -287.171 & -288.854 & 64 & 50.6075 & -10216.4 & -10747.6 & -10820.7 \\
    15 & 11.9377 & -324.603 & -338.769 & -340.719 & 65 & 51.2340 & -10582.4 & -11146.7 & -11226.6 \\
    16 & 12.7366 & -378.517 & -395.236 & -397.505 & 66 & 51.9530 & -10965.9 & -11557.2 & -11641.5 \\
    17 & 13.5314 & -437.400 & -456.884 & -459.482 & 67 & 52.6702 & -11357.4 & -11976.5 & -12065.3 \\
    18 & 14.3222 & -501.418 & -523.879 & -526.818 & 68 & 53.3856 & -11757.1 & -12404.7 & -12498.2 \\
    19 & 15.1910 & -571.305 & -595.918 & -599.165 & 69 & 54.0996 & -12165.2 & -12841.9 & -12940.2 \\
    20 & 16.0556 & -646.244 & -673.183 & -676.758 & 70 & 54.8124 & -12581.8 & -13288.3 & -13391.5 \\\cline{1-5}
    21 & 16.8063 & -723.779 & -755.341 & -759.736 & 71 & 55.6210 & -13017.6 & -13746.6 & -13851.8 \\
    22 & 17.5526 & -806.609 & -843.167 & -848.406 & 72 & 56.4286 & -13462.0 & -14213.9 & -14321.2 \\
    23 & 18.2939 & -894.773 & -936.754 & -942.884 & 73 & 57.2350 & -13915.1 & -14690.2 & -14799.8 \\
    24 & 18.9135 & -984.973 & -1035.65 & -1043.36 & 74 & 58.0403 & -14376.8 & -15175.6 & -15287.5 \\
    25 & 19.7636 & -1087.71 & -1141.81 & -1149.87 & 75 & 58.8447 & -14847.3 & -15670.3 & -15784.5 \\
    26 & 20.4882 & -1192.25 & -1253.26 & -1262.44 & 76 & 59.6472 & -15326.2 & -16173.9 & -16290.6 \\
    27 & 21.2099 & -1302.72 & -1371.06 & -1381.41 & 77 & 60.4487 & -15813.9 & -16686.8 & -16806.1 \\
    28 & 21.9279 & -1419.13 & -1495.28 & -1506.87 & 78 & 61.1879 & -16300.8 & -17208.2 & -17331.1 \\
    29 & 22.5146 & -1536.57 & -1625.17 & -1638.96 & 79 & 61.9874 & -16806.4 & -17739.9 & -17865.4 \\
    30 & 23.3548 & -1670.43 & -1763.60 & -1777.85 & 80 & 62.8473 & -17330.8 & -18281.7 & -18409.0 \\
    31 & 24.1826 & -1809.22 & -1908.47 & -1923.26 & 81 & 63.7020 & -17861.7 & -18832.7 & -18961.8 \\
    32 & 25.0083 & -1954.42 & -2060.00 & -2075.36 & 82 & 64.5560 & -18401.7 & -19393.0 & -19524.0 \\
    33 & 25.8319 & -2106.13 & -2218.27 & -2234.24 & 83 & 65.4092 & -18950.8 & -19962.7 & -20095.6 \\
    34 & 26.6516 & -2264.11 & -2383.19 & -2399.87 & 84 & 66.2610 & -19508.5 & -20541.5 & -20676.5 \\
    35 & 27.4694 & -2428.77 & -2555.03 & -2572.44 & 85 & 67.1120 & -20075.4 & -21129.8 & -21266.9 \\
    36 & 28.2853 & -2600.19 & -2733.88 & -2752.05 & 86 & 67.9623 & -20651.5 & -21727.6 & -21866.8 \\
    37 & 29.1585 & -2780.21 & -2919.52 & -2938.36 & 87 & 68.8470 & -21241.1 & -22334.9 & -22475.9 \\
    38 & 30.0296 & -2966.85 & -3111.99 & -3131.55 & 88 & 69.7309 & -21839.6 & -22951.5 & -23094.3 \\
    39 & 30.8213 & -3155.02 & -3310.59 & -3331.68 & 89 & 70.5707 & -22437.9 & -23576.6 & -23722.2 \\
    40 & 31.6110 & -3350.00 & -3516.37 & -3539.00 & 90 & 71.4096 & -23045.4 & -24211.2 & -24359.6 \\
    41 & 32.3199 & -3546.33 & -3728.64 & -3753.6 & 91 & 72.1161 & -23639.5 & -24850.7 & -25007.1 \\
    42 & 33.1052 & -3755.01 & -3949.05 & -3975.55 & 92 & 72.8875 & -24254.1 & -25502.5 & -25664.3 \\
    43 & 33.9674 & -3976.23 & -4177.47 & -4204.79 & 93 & 73.6578 & -24878.0 & -26164.2 & -26331.5 \\
    44 & 34.6664 & -4192.63 & -4411.72 & -4441.54 & 94 & 74.3604 & -25499.2 & -26833.3 & -27008.7 \\
    45 & 35.4442 & -4422.14 & -4654.37 & -4685.88 & 95 & 75.1286 & -26141.7 & -27515.0 & -27695.9 \\
    46 & 36.1379 & -4652.44 & -4903.88 & -4937.92 & 96 & 75.9626 & -26805.4 & -28209.0 & -28392.8 \\
    47 & 36.9945 & -4902.97 & -5162.74 & -5197.7 & 97 & 76.7280 & -27466.1 & -28910.3 & -29099.8 \\
    48 & 37.8493 & -5160.83 & -5429.17 & -5465.13 & 98 & 77.4251 & -28124.0 & -29619.4 & -29817.4 \\
    49 & 38.6966 & -5424.43 & -5703.19 & -5740.17 & 99 & 78.1888 & -28803.8 & -30341.2 & -30545.0 \\
    50 & 39.5426 & -5695.47 & -5984.91 & -6022.93 & 100 & 78.9514 & -29493.1 & -31073.2 & -31282.8 \\
    \hline \hline
\caption{An effective charge $Z^{*}$ and the comparison of the energy in a.u. of the zeroth-order approximation and the second-order perturbation theory in the hydrogen-like basis (single-particle excitation) Eq.~(\ref{eq:19}) with the values via HF \cite{Saito2009836}. The line after $Z = 20$ indicates that for larger $Z$ the inclusion of relativistic corrections is important \cite{HUANG1976243}.}
\label{tab:2}
\end{longtable*}

\bibliography{analytic_model_of_an_atom}

\end{document}